\documentclass[letterpaper]{article} % DO NOT CHANGE THIS
\usepackage{aaai24}  % DO NOT CHANGE THIS
\usepackage{times}  % DO NOT CHANGE THIS
\usepackage{helvet}  % DO NOT CHANGE THIS
\usepackage{courier}  % DO NOT CHANGE THIS
\usepackage[hyphens]{url}  % DO NOT CHANGE THIS
\usepackage{graphicx} % DO NOT CHANGE THIS
\usepackage{natbib}  % DO NOT CHANGE THIS AND DO NOT ADD ANY OPTIONS TO IT
\usepackage{caption} % DO NOT CHANGE THIS AND DO NOT ADD ANY OPTIONS TO IT
\usepackage{algorithm}
\usepackage{algorithmic}

\usepackage{newfloat}
\usepackage{listings}
\DeclareCaptionStyle{ruled}{labelfont=normalfont,labelsep=colon,strut=off} % DO NOT CHANGE THIS
\floatstyle{ruled}
\newfloat{listing}{tb}{lst}{}
\floatname{listing}{Listing}
\usepackage{xcolor}
\usepackage{amsfonts} 
\usepackage{subcaption}
\usepackage{rotating}

\newcommand{\citenoun}[1]{{\citet{#1}}}
\usepackage{xcolor}
\newcommand{\answerYes}[1]{\textcolor{blue}{#1}} 
\newcommand{\answerNo}[1]{\textcolor{teal}{#1}} 
\newcommand{\answerNA}[1]{\textcolor{gray}{#1}}

\usepackage{amsmath}
\usepackage{multirow}

\usepackage{enumerate}
\usepackage{enumitem}

\title{Forecasting Political News Engagement on Social Media}
\author {
    Karthik Shivaram\textsuperscript{\rm 1},
    Mustafa Bilgic\textsuperscript{\rm 2}, 
    Matthew Shapiro\textsuperscript{\rm 3}, 
    Aron Culotta\textsuperscript{\rm 1}
}
\affiliations {
    \textsuperscript{\rm 1} Department of Computer Science, Tulane University, New Orleans, LA  ~ USA\\
    \textsuperscript{\rm 2} Department of Computer Science, Illinois Institute of Technology, Chicago, IL  ~ USA\\
    \textsuperscript{\rm 3} Department of Social Sciences, Illinois Institute of Technology, Chicago, IL  ~ USA\\
    kshvaram@tulane.edu, mbilgic@iit.edu, shapiro@iit.edu, aculotta@tulane.edu
}
\begin{document}

\maketitle

\begin{abstract}
Understanding how political news consumption changes over time can provide insights into issues such as hyperpartisanship, filter bubbles, and misinformation. To investigate long-term trends of news consumption, we curate a collection of over 60M tweets from politically engaged users over seven years, annotating $\sim$10\% with mentions of news outlets and their political leaning. We then train a neural network to forecast the political lean of news articles Twitter users will engage with, considering both past news engagements as well as tweet content. Using the learned representation of this model, we cluster users to discover salient patterns of long-term news engagement. Our findings include the following: (1) hyperpartisan users are more engaged with news; (2) right-leaning users engage with contra-partisan sources more than left-leaning users; (3) topics such as immigration, COVID-19, Islamaphobia, and gun control are  salient indicators of engagement with low quality news sources.
\end{abstract}

\section{Introduction}

The transformation of the news ecosystem from print to online media has fundamentally changed how people read about and engage with current events. While this media decentralization has undoubtedly increased one's access to diverse and timely information, it has also led to hyperpartisanship, polarization, and misinformation, all of which are fostered by computer-mediated communication. Recent work has investigated socio-technical issues such as the causes of polarization and ``filter bubbles''~\cite{bakshy2015exposure,bail2018exposure,dellaposta2020pluralistic,liu2021interaction,shivaram2022reducing}, the influence of hyperpartisan media~\cite{king2017news,guess2021consequences}, and the factors that contribute to the spread of misinformation~\cite{lobato2020factors,osmundsen2021partysharenews}. It has also led to research focusing on identifying interventions that can help foster healthier news engagement habits~\cite{bhargava2019gobo,lutzke2019priming,masrour2020bursting}.

Among such scholarship, however, studies have typically been either small controlled trials or observational studies over very short time periods. Those that consider long-term trends are primarily focusing on population-level changes over time. What is lacking is a long-term perspective to detail the evolution of an individual's news consumption and sharing habits. In response, we collect over 60M tweets from politically engaged Twitter users over a seven year period, and we annotate $\sim$10\% with mentions of news outlets and the political lean of those outlets. With this data, we investigate how the types of news sources a user engages with change over long time periods. Our primary contributions are as follows:
\begin{itemize}
\item\textbf{Novel Dataset:} We curate a new dataset to foster research into long-term news engagement behavior, consisting of $\sim$6.5M news engagement tweets over seven years. To preserve anonymity, the dataset consists of records of the mentioned news source, partisan lean, day, and anonymized user IDs.\footnote{Code and data are available at: \url{https://github.com/tapilab/icwsm-2024-news-forecasting}}
\item\textbf{Forecasting Model:} We develop LSTM models to forecast news engagement behavior over three-month windows, predicting the number of news mentions from each of seven political stance categories (extreme liberal to extreme conservative). The best performing model has a mean absolute error of 3.7 engagements (out of an average of $\sim$39 news engagements per window).
\item \textbf{Discovery of Long-term Patterns:} The intermediate states of the LSTM offer a compact representation of the long-term behavior of each user. We cluster users to discover salient patterns, finding that (1) hyperpartisan users are more engaged with the news; (2) right-leaning users engage with contra-partisan sources more than left-leaning users do; (3) topics such as immigration, COVID-19, Islamaphobia, and gun control are leading indicators of engagement with low quality news sources.
\end{itemize}

\section{Related Work}

One area of relevant research considers the factors contributing to engagement with hyperpartisan news, misinformation, and conspiracy theories. In some cases, the content is highlighted, such as its topic and tone~\cite{wischnewski2021shareworthiness}, as well as how such content aligns with the political views of the user. For example, \citenoun{osmundsen2021partysharenews} show that hostility towards political opponents drives misinformation sharing, \citenoun{rathje2021out} find that such out-group language strongly predicts social media engagement, and \citenoun{brady_etal2017} and \citenoun{valenzuela2017} find that moral-emotional language in political messages increases their diffusion. Other related research focuses on the personality traits of the user; e.g., \citenoun{lobato2020factors} show that individuals with personality traits high in traditionalism and low in social dominance were more willing to share misinformation about COVID-19, and a meta-analysis of COVID-19 misinformation by \citenoun{van2020antecedents} finds that biases, group identity, and distrust in institutions contribute to misinformation sharing. 
There are also investigations into how hyperpartisan news and misinformation spreads online~\cite{haber2021research,introne2020mapping}, as well as possible interventions for minimizing users' engagement with such content~\cite{bak2022combining,bhuiyan2021nudgecred,masrour2020bursting,pennycook2021shiftingattention, aslett2022,nyhan2021}. Yet, beyond an examination of how users navigate YouTube to access more extreme content~\cite{ribeiro2020auditing}, there is little research on user-level long-term trends in news engagement.

A second line of research investigates the dynamics surrounding social media-based filter bubbles, polarization, and ideological segregation. Much has been written on these topics, and we highlight a selection of the most relevant findings, beginning with \citenoun{dellaposta2020pluralistic}, who shows that polarization arises not simply by hardening opinions on a handful of issues but rather by belief consolidation, in which partisan views on one topic spread to other topics. \citenoun{robertson2021engagement} provide evidence that filter bubbles are driven less by the idiosyncrasies of search engines and more by users self-selecting into ``echo chambers.'' This is consistent with the finding of \citenoun{bakshy2015exposure} that polarization on Facebook is mostly driven by homophily of user friendship networks. Relatedly, \citenoun{bail2018exposure} show that exposure to opposing views on social media can increase political polarization. Other work has performed linguistic analysis to track how issues are framed differently by news source and over time~\cite{tsur2015frame,liu2022climate,islam2023weakly}.

Much of this prior work focuses on correlations among static user attributes and news sharing behavior, or it conducts randomized controlled trials that investigate a handful of interventions. Given the complex factors influencing news engagement over the short- and medium-terms, we call for new methods to analyze longitudinal, real-world data and explore patterns of news engagement behavior over time. We propose several neural network architectures and learning methods to more accurately predict future news sharing behavior of users as well as to identify latent representations characterizing a user's transitioning news engagement.
 
\section{Data}

\begin{table}[t]
\small
\centering
\begin{tabular}{crrr}
\hline
\textbf{Year}  & \textbf{\begin{tabular}[c]{@{}c@{}}All \\ Tweets\end{tabular}} & \textbf{\begin{tabular}[c]{@{}c@{}}NE \\ Tweets\end{tabular}} & \textbf{NE \%} \\ \hline
2006           & 3                                                              & 0                                                                  & 0.00               \\
2007           & 424                                                            & 0                                                                  & 0.00               \\
2008           & 14,649                                                         & 7                                                                  & 0.05               \\
2009           & 92,788                                                         & 189                                                                & 0.20               \\
2010           & 187,609                                                        & 2398                                                               & 1.28               \\
2011           & 741,038                                                        & 16,065                                                             & 2.17               \\
2012           & 1,569,018                                                      & 49,765                                                             & 3.17               \\
2013           & 1,853,832                                                      & 74,069                                                             & 4.00               \\
2014           & 2,172,500                                                      & 120,568                                                            & 5.55               \\
2015           & 2,579,583                                                      & 181,957                                                            & 7.05               \\
2016           & 3,502,755                                                      & 350,700                                                            & 10.01              \\
2017           & 4,816,301                                                      & 585,272                                                            & 12.15              \\
2018           & 5,744,817                                                      & 684,334                                                            & 11.91              \\
2019           & 7,889,222                                                      & 904,271                                                            & 11.46              \\
2020           & 16,247,000                                                     & 1,843,002                                                          & 11.34              \\
2021           & 16,021,606                                                     & 1,654,322                                                          & 10.33              \\ \hline
\textbf{Total} & \textbf{63,433,145}                                            & \textbf{6,466,919}                                                 & \textbf{10.20}     \\ \hline
\end{tabular}
\caption{Tweets collected by year (NE = news engagement).}
\label{tab:tweet_dist_year_ex}
\end{table}

Our dataset is based on tweets where users engage, over a period of several years, with news sources across different partisan and ideological dimensions. For the present study, we focus on ``news engagement'' events where a user either mentions the official Twitter handle of a news source or shares a URL to an article from that source. We first identify a set of English language-based news sources covering the partisan landscape. We then assign political ideology ratings to different news sources based on \textit{allsides.com}, a media bias rating site, where the ideology spectrum \{-2,-1,0,+1,+2\} represents -2 for extreme liberals and +2 for extreme conservatives. We utilize ratings for 419 different news sources; yet, to account for a more diverse range of media quality, we extend the sample to include 103 low-reliability sources collected from \citet{osmundsen2021partisan},\footnote{We collected data for 167 news sources from \citet{osmundsen2021partisan}, excluding those that were either (a) already included in AllSides or (b) without a Twitter account or website. This reduced this subsample to 103 sources.} which were originally identified by \citenoun{guess2019fake} and \citenoun{grinberg2019fake}. These low-reliability sources are rated as being either \textit{pro-Republican} or \textit{pro-Democrat}, and we assign the partisan lean of these sources as -3 (pro-Democrat) and +3 (pro-Republican). The final dataset is thus comprised of 522 news sources that have valid Twitter handles and URLs.

We next use the Twitter Streaming API to identify users who engage with each of the above mentioned 522 news sources. %``Engagement'' refers to those instances when a user either mention's the Twitter handle of the news source or shares a URL that matches its web domain. 
We submitted queries to the API in Fall 2021 to identify mentions of each news source, resulting in the identification of 1.67 million users. To account for the diversity of all Twitter users (i.e., users who do not engage with any of the identified news sources), we also collected data for 59K random users using the Twitter API over the same period. We limited the presence of bots by filtering out users that appear to exhibit automated behavior based on frequency of tweets, number of followers, and number of friends (c.f., Appendix). Based on the news sources with which they engage, we then sampled $\sim$1,200 users by partisan stance, evenly distributed across news sources per stance. We combined this with 1,200 random users sampled from the Twitter API, resulting in an initial set of 9,781 users.

We collected each sampled user's entire Twitter timeline, amounting to a total of 63.5 million tweets. The annual tweet distribution for news engagement tweets is shown in Table \ref{tab:tweet_dist_year_ex}, where ``NE Tweets'' refer to those instances when a user engages with a given news source, encompassing 10.2\% of the tweets collected.  The distribution of tweets by partisan stance can be seen in Table \ref{tab:tweet_dist_ps_ex},  where the most commonly engaged news sources are moderate liberal/Democrat (-1), followed by moderate/non-partisan (0) and strongly conservative/Republican (+2).

\begin{table}[t]
\small
\centering
\begin{tabular}{rrr}
\hline
                & \multicolumn{2}{c}{\textbf{News Engagement Tweets}} \\
\textbf{Stance} & \multicolumn{1}{c}{Number} & {Percentage} \\
\hline
Unreliable Liberal (-3)              & 112,560                                                             & 1.74               \\
Extreme Liberal (-2)              & 1,141,939                                                            & 17.66              \\
Liberal (-1)              & 1,977,177                                                            & 30.57              \\
Non-partisan (0)               & 1,240,848                                                            & 19.19              \\
Conservative (1)               & 569,489                                                             & 8.81               \\
Extreme Conservative (2)               & 1,256,521                                                            & 19.43              \\
Unreliable Conservative (3)               & 168,385                                                             & 2.60               \\ \hline
\textbf{Total}  & \textbf{6,466,919}                                                 & \textbf{100}        \\ \hline
\end{tabular}
\caption{News engagement tweets by partisan stance.}
\label{tab:tweet_dist_ps_ex}
\end{table}

Table \ref{tab:tweet_dist_year_ex} shows that the majority of tweets in our dataset were posted after 2014, and we thus analyze tweets made after January 1, 2015. To focus on users with a significant history of news engagement, we exclude user accounts that do not engage with news sources at least 50 times since 2015. We also exclude users whose news engagement is not distributed across at least a three-year period. Based on these filtering criteria, 3,806 users were removed from our sample. We also 
filter out 137 potentially automated accounts in our sample by removing users whose total news engagement volume is three standard deviations greater than the mean news engagement volume (i.e., 3-sigma rule) \cite{hasan2018survey}. In sum, our final dataset is comprised of 5,838 users who generate 44.2M total tweets, of which 4.6M are news engagements.

\section{Problem Statement}

We forecast a user's future news engagement behavior based on their prior activity. To formulate this task, for user $u_i$ in time window $t_j$, let $y_{t_j}^{i} \in \mathbb{N}^{1 \times 7}$ be a vector representing the number of news engagements with each of the seven partisan stances $\{-3,-2,-1,0,1,2,3\}$. As described earlier, a news engagement occurs when a user either mentions the Twitter handle or shares a URL to an article from a news source.

As an input to the model, let $m_{t_j}^{i}$ represent attributes of all tweets (text, mentions, hashtags, etc.) of user $u_i$ across $t_j$. Given the observed historic news engagement count vectors \{$y_{t_1}^{i}, y_{t_2}^{i}, ..., y_{t_n}^{i}$\} from time window $t_1$ to $t_n$ and tweet attributes \{$m_{t_1}^{i},m_{t_2}^{i},...,m_{t_n}^{i}$\}, our goal is to estimate the probability distribution over the engagement counts of a future time step $t_{n+1}$ for user $u_i$.

\begin{equation}
    P(y_{t_{n+1}}^{i} | y_{t_1}^{i}...y_{t_n}^{i}, m_{t_1}^{i}...m_{t_n}^{i})
\end{equation}
In other words, given historic tweet attributes and engagement counts, we intend to predict the number of times a given user engages with news sources across all stances ($-3$ to $+3$) in subsequent time steps.

\section{Methods}

We discuss in this section our data pre-processing techniques, a number of forecasting models employed via multiple neural network architectures, and our baseline method. We also describe the different features that we employ for training our models.

\begin{table}[t]
\small
\centering
\begin{tabular}{cccc}
\hline
\textbf{Dataset} & \textbf{Train} & \textbf{Val} & \textbf{Test} \\ \hline
$D_1$            & 2015-2017      & 2015-2017    & 2018          \\
$D_2$            & 2016-2018      & 2016-2018    & 2019          \\
$D_3$            & 2017-2019      & 2017-2019    & 2020          \\
$D_4$            & 2018-2020      & 2018-2020    & 2021          \\ \hline
\end{tabular}
\caption{Time ranges for each train/test split of the forecasting task.}
\label{tab:data_division}
\end{table}

\begin{figure}[t]
    \centering
    \includegraphics[width=1.0\linewidth]{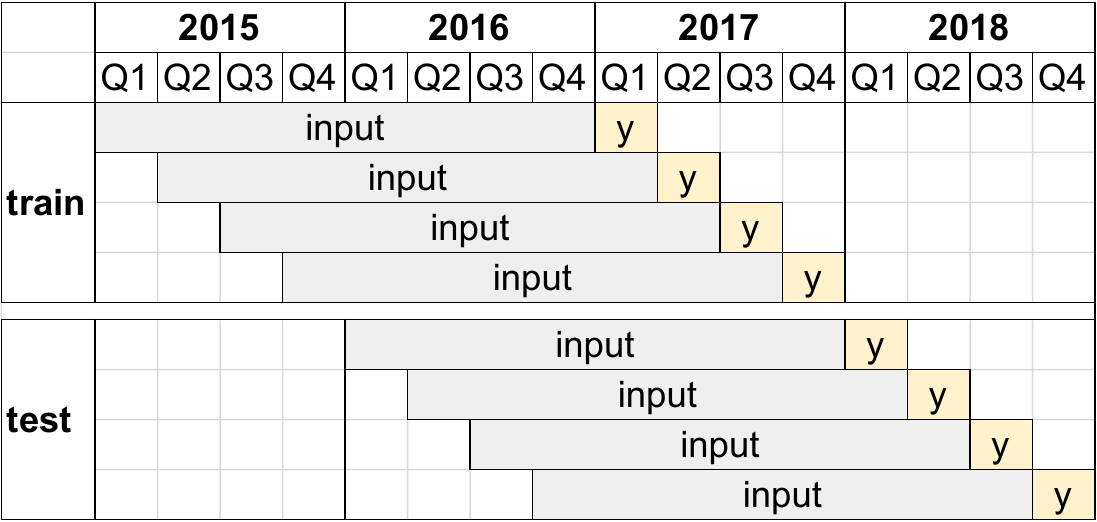}
    \caption{Train-test splits for $D_1$ (2015-2018), showing the time ranges used for each instance. Analogous time ranges are used for $D_2$-$D_4$.}
    \label{fig.splits}
\end{figure}

\subsection{Pre-processing}

We set $n=8$ (the number of input time steps), where each time step $t_j$ spans three months. Thus, given a user's activity over the previous two years, we predict the next three months of news engagements. To create multiple train/test folds, we split our overall dataset ($D$) into four folds \{$D_1,D_2,D_3,D_4$\} shown in Table~\ref{tab:data_division}, each of which spans three years. Figure~\ref{fig.splits} illustrates the input and output for the training and testing instances of $D_1$. Thus, for each user of $D_1$, we generate four training instances and four testing instances. This is repeated for each fold $D_1$ through $D_4$.

\subsection{Features}

\begin{enumerate}
    \item \textbf{News Engagement Counts} ($y_{t_j}^{i}$): For each input time step $t_j$ and user $u_i$, we include as input features the lagged count vectors of the number of engagements for each partisan stance $p \in \{-3,-2,-1,0,1,2,3\}$. We denote this $y_{t_j}^{i}$, where $y_{t_j}^{i} \in \mathbb{R}^{1 \times 7}$. We also standardize these count values using z-score standardization, where the means and standard deviations are calculated over the input time steps (i.e., $t_1$ to $t_n$).
    
    \item \textbf{Tweet Text} ($v_{t_j}^{i}$): We also include features over the tweet text. For each three month time step, we select the 25 most recent tweets for both the news engagement tweets and the non-news engagement tweets. We then pass these two sets of tweets through TwHIN-Bert \cite{zhang2022twhin}, a transformer based language model fine-tuned on Twitter data, and extract embedding representations for each token of each tweet. We perform two levels of aggregation over these token representations: (1) for each tweet, we concatenate the CLS token and average embedding of the other non-CLS tokens of the tweet; (2) we take an average over these 25 tweet representations. This results in two 1,536-dimension vector representations, one for the engagement tweets ($\text{eng}_{t_j}$) and one for the non-engagement tweets ($\text{neng}_{t_j}$). We then concatenate these two embedding vectors to get a single text representation $v_{t_j}^{i}$, where $v_{t_j}^{i} \in \mathbb{R}^{1 \times 3072}$.

    \item \textbf{Hashtags} ($\#^{i}_{t_n}$): For the last input time step $t_n$, we select the top 100 most frequently used hashtags for that specific three-month window. We then pass these hashtags through the language model (TwHIN-Bert) and perform the same aggregation step that we used for our text-based features, obtaining a final vector representation $\#^{i}_{t_n}$ where $\#^{i}_{t_n} \in \mathbb{R}^{1 \times 1536}$.

    \item \textbf{Input Quarter Encoding} ($q^{i}_{t_j}$): To capture seasonal effects, for each observation sequence we encode the year-quarter of each input time step (Q1 - Q4) as a one-hot encoded vector $q^{i}_{t_j}$, where $q^{i}_{t_j} \in \mathbb{R}^{1 \times 4}$. 

    \item \textbf{Forecast Quarter Encoding}: Similarly, for each observation sequence, we encode the year-quarter of the time-step we are forecasting as a one-hot encoding vector.
\end{enumerate}

\subsection{Baseline}

For comparison, we include a simple approach that sets the prediction to be equal to the values for the final input time step (i.e., $\hat{y}^{i}_{t_{n+1}} =  y_{t_n}^i$). 

\subsection{Single Feature Network (SFN)}

We use a bi-directional LSTM (Bi-LSTM) \cite{schuster1997bidirectional} as our primary forecasting model. We select this model in part due to its application in other types of user activity forecasting -- e.g.,  \citet{yang2018know} fit an LSTM to predict churn rate for a social media application, and \citet{liu2019characterizing} used GNNs to forecast future user engagement on Snapchat.
Similar models have performed best on time series benchmark competitions~\cite{oreshkin2019n}.\footnote{While transformer-based models are a viable alternative, recent work has questioned the effectiveness of such models for time series forecasting~\cite{zeng2023transformers}.} 

The Single Feature Network (SFN) uses only one type of feature --- either text based representation sequences \{$v_{t_1}^{i}, v_{t_2}^{i}, ..., v_{t_n}^{i}$\} (\textbf{SFN+T}) or news engagement count based sequences \{$y_{t_1}^{i}, y_{t_2}^{i}, ..., y_{t_n}^{i}$\} (\textbf{SFN+C}). When using the text based representations we add a linear layer before passing the input sequences into the Bi-LSTM. After passing the input sequences through our Bi-LSTM model, we extract the final hidden states for both the forward and backward layers ($\overrightarrow{h_{t_n}}, \overleftarrow{h_{t_n}}$) and concatenate them to obtain a single hidden state representation $h_{t_n}$. This is then passed through a final output layer \ \textless$W_{out},b_{out}$\textgreater \ to predict the future news engagement count vector $\hat{y}^i_{t_{n+1}}$ for time step $t_{n+1}$, as shown in Equation \ref{equ:bi_lstm_sfn_out}.

\begin{equation}\label{equ:bi_lstm_sfn_out}
    \hat{y}^i_{t_{n+1}} = (W_{out}h_{t_n} + b_{out})
\end{equation}

\subsection{Multiple Feature Network (MFN)}

The Multiple Feature Network (MFN) combines multiple features (tweet texts, news engagement counts, and input quarter encodings) to forecast future news engagement counts. This model's architecture resembles the single feature network (SFN) with a few modifications. Once we extract the final hidden state representation $h_{t_n}$ from our Bi-LSTM layers (discussed above), we concatenate the hashtag representation $\#^{i}_{t_n}$ of the final input time step and the output quarter encoding $q_{t_{n+1}}^i$ to this hidden state representation $h_{t_n}$. This is then passed through an intermediate layer \ \textless$W_{inter},b_{inter}$\textgreater \ , the output of which, $g^{i}_{inter}$, is then passed through our final output layer \ \textless$W_{out},b_{out}$\textgreater \ to predict the future news engagement count vector $\hat{y}^i_{t_{n+1}}$ for time step $t_{n+1}$, as shown in equation \ref{equ:bi_lstm_mfn_out}.

\begin{equation}
    g^{i}_{inter} = (W_{inter}[h_{t_n}, \#^{i}_{t_n}, q_{t_{n+1}}] + b_{inter})
\end{equation}

\begin{equation}\label{equ:bi_lstm_mfn_out}
    \hat{y}^i_{t_{n+1}} = (W_{out} g^{i}_{inter} + b_{out})
\end{equation}

All network models are trained to minimize Mean Absolute Error (MAE) loss.\footnote{Other loss functions such as Mean Squared Error (MSE), Mean Absolute Percentage Error (MAPE), and Huber Loss were also considered, but exploratory experiments suggested MAE was less sensitive to outliers.}  As we predict a vector of engagement counts, the overall MAE loss is a sum across individual MAE losses for each news engagement stance:

\begin{equation} \label{equ:mae_loss_total}
    \text{Total MAE Loss} = \sum_{r=1}^{|p|} \text{MAE}(y_{t_{n+1}}[r], \hat{y}_{t_{n+1}}[r])
\end{equation}

\section{Experimental Settings}

To perform our forecasting experiments, we first construct our train, test, and validation sets for each fold $D_1$ through $D_4$ (Table \ref{tab:data_division}). For validation, we hold out 20\% of the users in the training set.  Once we have our train, validation, and test sets, we filter out sequences where a user has no news engagement activity over the entire input sequence. The resulting dataset sizes are shown in Table \ref{tab:train_val_test_sizes_forecast}.

All experiments use a system with 4 Nvidia A5000 GPUs, 512 GB RAM and an AMD Ryzen Threadripper 3975WX CPU. We report means and standard deviations of scores across five random seeds. Table~\ref{tab:hyperparameters} in the Appendix lists all hyperparameters that were tuned on the validation data.

\begin{table}[t]
\centering
\begin{tabular}{cccc}
\hline
\textbf{Dataset} & \textbf{Train} & \textbf{Val} & \textbf{Test} \\ \hline
$D_1$               & 15,708         & 3,904         & 21,881        \\
$D_2$               & 17,536         & 4,345         & 22,327        \\
$D_3$               & 17,862         & 4,465         & 22,648        \\
$D_4$               & 18,106         & 4,542         & 17,378        \\ \hline
\end{tabular}
\caption{Train, validation, and test set sizes. Each instance is a single user's timeline, with two years of input observations used to forecast the next three months of news engagements.}
\label{tab:train_val_test_sizes_forecast}
\end{table}

\section{Results and Analysis}
We now turn to the results of our experiments and conduct an extended analysis of long-term news engagement behavior of users on Twitter.

\begin{table}[t]
\centering
\begin{tabular}{ccc}
\hline
\textbf{Model} & \textbf{MAE} & \textbf{MSE} \\ \hline
Baseline       & 3.89            & 216.33          \\
SFN+C          & \textbf{3.73 }           & \textbf{207.89}          \\
SFN+T          & 4.22            & 258.08          \\
MFN            & 3.85            & 220.55          \\ \hline
\end{tabular}
\caption{Mean absolute error and mean-squared error averaged over all test folds, showing SFN+C as the most accurate model.}
\label{tab:agg_metrics_forecast}
\end{table}

\begin{table*}[t]
\small
\centering
\begin{tabular}{ccccccccc}
\hline
\textbf{Dataset}    & \textbf{Model} & \textbf{-3}    & \textbf{-2}          & \textbf{-1}           & \textbf{0}           & \textbf{1}           & \textbf{2}           & \textbf{3}                   \\ \hline
\multirow{4}{*}{$D_1$} & Baseline       & \textbf{.20~~~~~~~~~} & 3.14~~~~~~~~~                & 5.09~~~~~~~~~                 & 3.21~~~~~~~~~                & 1.32~~~~~~~~~                & 3.08~~~~~~~~~                & \textbf{.49~~~~~~~~~}                      \\
                    & SFN+C          & .21±.005     & \textbf{3.00±.004} & \textbf{4.80±.012}  & \textbf{2.98±.008} & \textbf{1.29±.007} & \textbf{2.93±.003} & .54±.012           \\
                    & SFN+T          & .27±.004    & 3.29±.031          & 5.40±.045             & 3.36±.043          & 1.40±.014            & 3.30±.053          & .57±.021                    \\
                    & MFN            & .27±.001    & 3.08±.029          & 4.97±.116           & 3.09±.076          & 1.34±.025          & 3.05±.046          & .57±.028                   \\ \hline
\multirow{4}{*}{$D_2$} & Baseline    & \textbf{.36~~~~~~~~~} & 3.70~~~~~~~~~                & 6.24~~~~~~~~~                 & 3.80~~~~~~~~~                & 2.01~~~~~~~~~                & 3.10~~~~~~~~~                & .45~~~~~~~~~                                \\
                    & SFN+C          & .39±.042    & \textbf{3.46±.112} & \textbf{5.88±.335}  & \textbf{3.55±.132} & \textbf{1.90±.155} & \textbf{2.99±.082} & \textbf{.43±.008}   \\
                    & SFN+T          & .40±.001    & 3.91±.015          & 6.78±.025           & 3.98±.007           & 2.18±.011          & 3.50±.049          & .52±.016                    \\
                    & MFN            & .41±.0      & 3.55±.018          & 6.10±.024           & 3.64±.010           & 1.99±.007          & 3.07±.034          & .44±.005                   \\ \hline
\multirow{4}{*}{$D_3$} & Baseline       & \textbf{.62~~~~~~~~~} & 6.39~~~~~~~~~                & 10.84~~~~~~~~~                & 6.98~~~~~~~~~                & 3.57~~~~~~~~~                & \textbf{6.51}~~~~~~~~~                & \textbf{.68~~~~~~~~~}                      \\
                    & SFN+C          & .64±.007    & \textbf{6.01±.029}  & \textbf{10.46±074} & 6.74±.053          & \textbf{3.49±.014} & \textbf{6.51±.065} & .70±.002           \\
                    & SFN+T          & .69±.001    & 6.55±.019          & 11.40±.035          & 7.27±.023          & 3.82±.010           & 7.34±.114          & .77±.003                   \\
                    & MFN            & .69±.001    & 6.06±.020           & 10.53±.111          & \textbf{6.73±032} & 3.49±021          & 6.58±.058          & .73±.004                    \\ \hline
\multirow{4}{*}{$D_4$} & Baseline       & \textbf{1.39~~~~~~~~~} & 6.07~~~~~~~~~                & 9.97~~~~~~~~~                 & 7.63~~~~~~~~~                & 4.92~~~~~~~~~                & 8.40~~~~~~~~~                & \textbf{1.23~~~~~~~~~}                       \\
                    & SFN+C          & 1.44±.017    & \textbf{5.73±.099} & \textbf{9.77±.317}  & \textbf{7.32±.043} & 4.75±.040           & \textbf{8.31±.137} & 1.35±.026  \\
                    & SFN+T          & 1.50±.005    & 6.67±.073          & 11.69±.473          & 8.33±.147          & 5.04±.037          & 9.02±.157          & 1.40±.003                   \\
                    & MFN            & 1.52±.002    & 6.04±.097           & 10.51±.185          & 7.89±.259          & \textbf{4.73±.041} & 8.57±.152          & 1.41±.011                     \\ \hline
\end{tabular}
\caption{Mean absolute error by partisan stance (smallest errors in bold).}
  \label{tab:performance_by_stance}
\end{table*}

\subsection{Model Accuracy}

Table~\ref{tab:agg_metrics_forecast} shows the average error rates of the models across all test datasets, $D_1$ through $D_4$, averaged across stance and dataset. We find that SFN+C and MFN perform first- and second-best, respectively, both outperforming the baseline model. Even though the SFN+T model performs the worst, we are surprised to discover that it performs relatively well for a model using text features without news engagement counts. To assess these differences in performance, we conduct paired t-tests over the combined results and find that, with the exception of the Baseline-MFN comparison, all pairwise comparisons yield significantly different average error rates (\textit{p} $<$ 0.01).

Table \ref{tab:performance_by_stance} breaks down error rates by fold and partisan stance. We can see that SFN+C has the lowest error for most cases. Exceptions include -3/+3, where the LSTM models generally do not outperform the baseline. We attribute this in part to class imbalance --- the overall infrequency of -3/+3 stances (c.f., Table~\ref{tab:tweet_dist_ps_ex}) lead to under prediction of engagement.

\begin{figure}[t]
    \centering
    \includegraphics[width=1.0\linewidth]{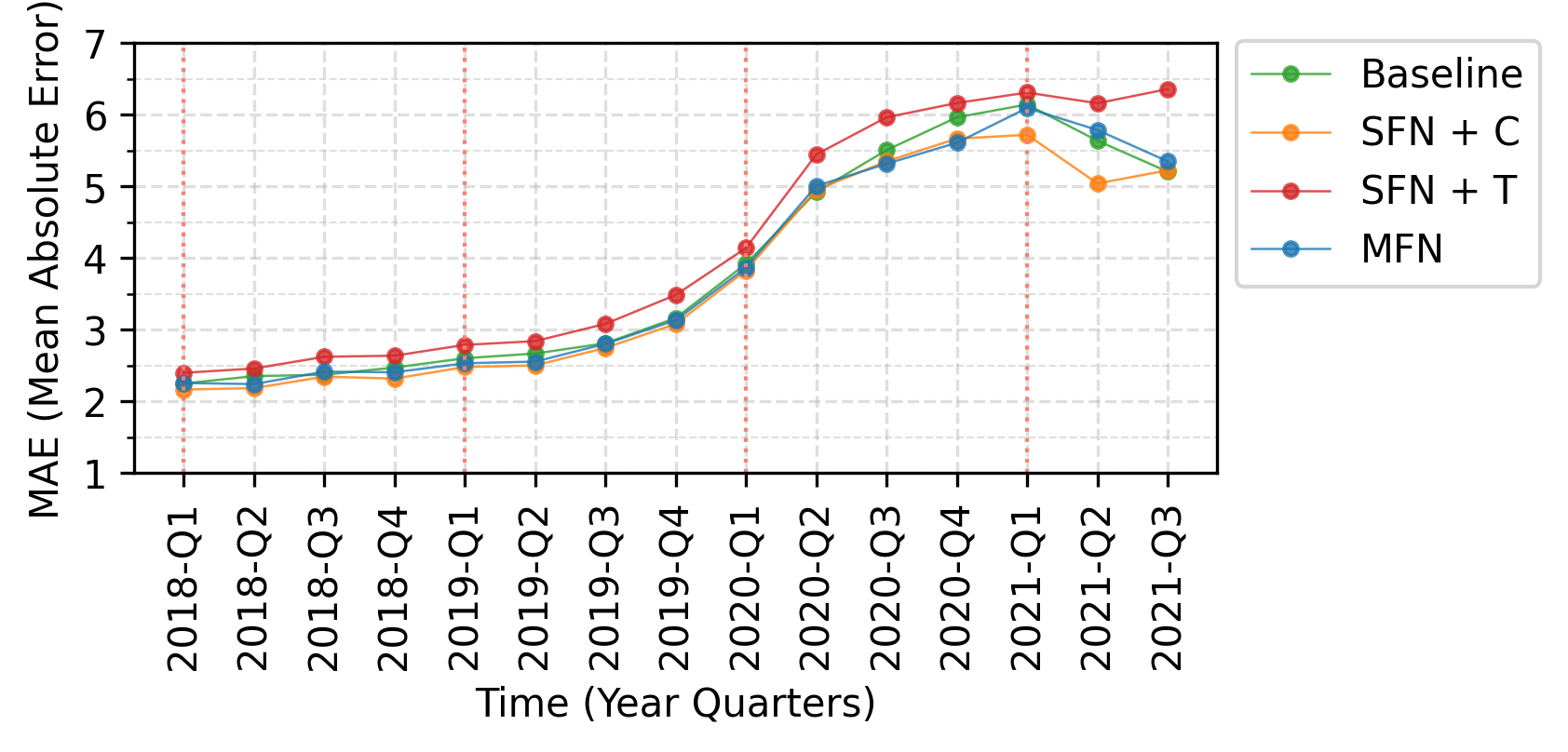}
    \caption{Mean absolute error by quarter. The rising error is in part due to increased user activity since 2020.}
    \label{fig.mae_by_quarter}
\end{figure}

Figure \ref{fig.mae_by_quarter} plots test error by quarter. We observe that error magnitude generally increases over time due largely to the overall increase in engagement volume. The highest increase in errors is measured for 2020, possibly resulting from a number of events that increased news engagement, namely, COVID-19 and Trump's impeachment. Over time, the SFN+C model performs best, with considerable improvements over other models from 2020-Q4 to 2021-Q2.

\subsection{User-level Error Analysis}

\begin{figure*}[ht]
    \centering
    \includegraphics[width=0.97\linewidth]{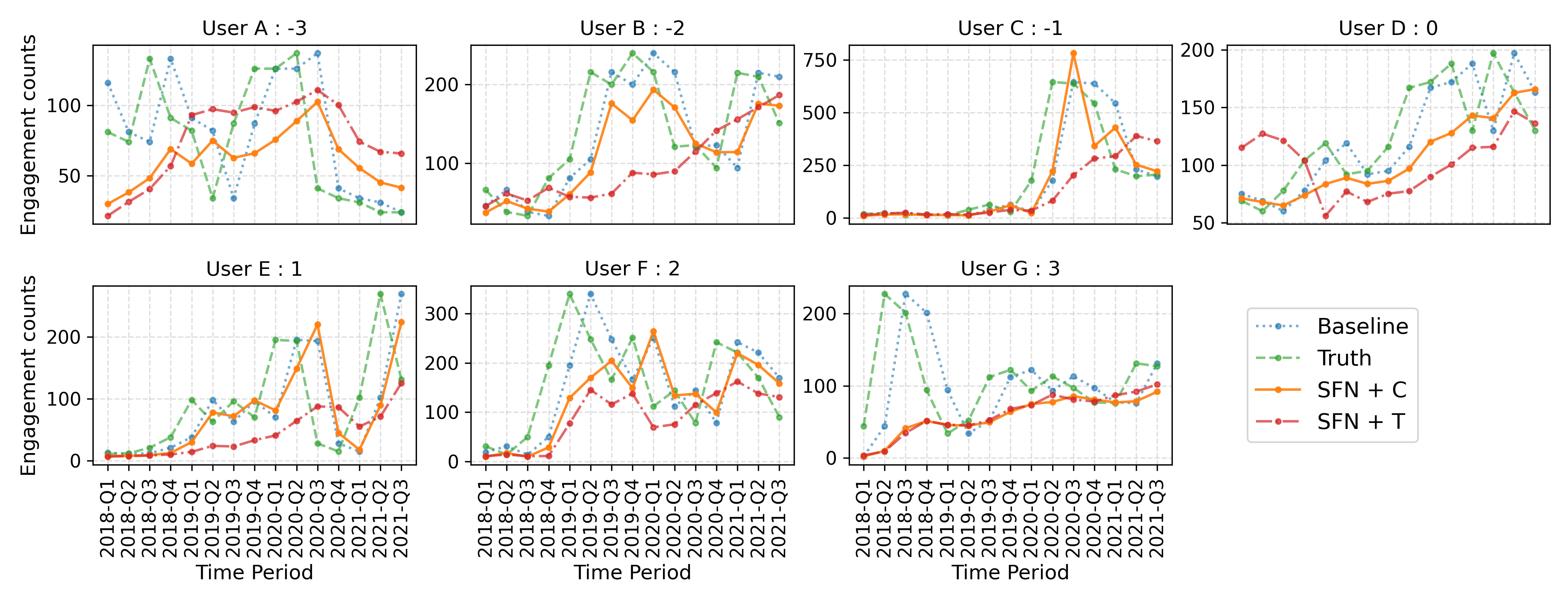}
    \caption{True vs. predictied values for users with the highest errors, according to the SFN+C model.}
    \label{fig.user_errors}
\end{figure*}

To gauge model performance for individual users, we plot the true and predicted counts for a sample of users in Figure~\ref{fig.user_errors}. These users were selected because they exhibited the highest error rates according to the SFN+C model.
Each subplot in Figure \ref{fig.user_errors} represents a different user-stance combination. For most stances, we observe that our models perform well when accounting for the overall trend of engagement counts. The main challenge for these models are sudden spikes in engagement. Yet, the SFN+C model seems to handle engagement spikes better than the SFN+T model, likely due to the fact that text based features do not capture the intensities of the news engagement for these users. We also observe that neither the SFN+C nor SFN+T models perform very well for the -3 category, which could result from the low volume of these types of engagements in our sample.

\subsection{Predicting Sudden Shifts in News Engagement}

\begin{figure}[t]
    \centering
    \includegraphics[width=0.8\linewidth]{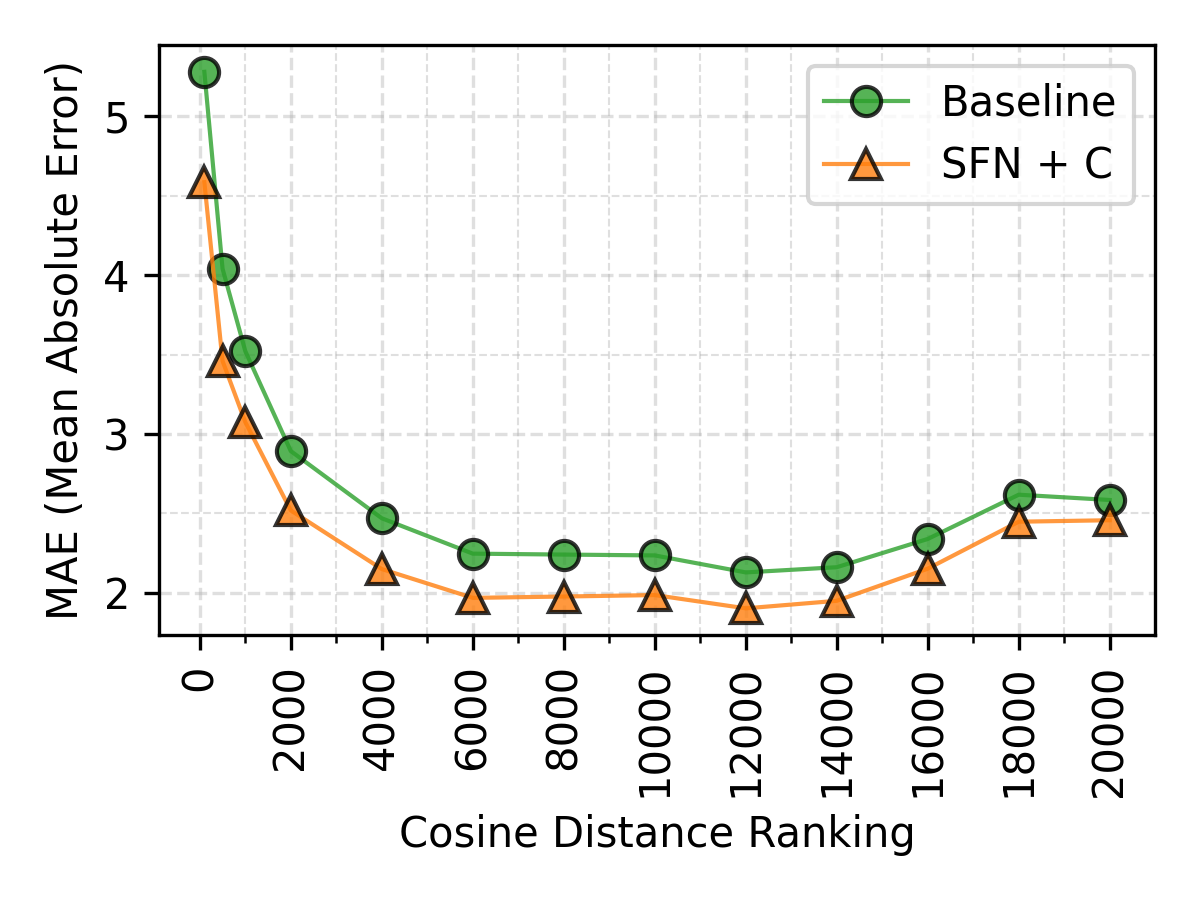}
    \caption{Error by transition points based on cosine distance ranking (lower ranking means more abrupt transitions).}
    \label{fig.hardness_plots}
\end{figure}

Sudden shifts in engagement may be difficult to predict, but they are of practical interest given that they represent a user's news engagement transition point. To examine this phenomenon more closely, we compare model performance across a range of transition points. To identify transitions where there is a shift in engagement between the input time steps (i.e., $t_{1}$ to $t_{n}$) and the forecast window ($t_{n+1}$), we rank samples by measuring the cosine distance between the engagement count vector of the last time-step of the input sequence ($y^i_{t_{n}}$) and the count vector to be predicted ($\hat{y}^i_{t_{n+1}}$). We next compute error metrics at all ranks of these %difficulty levels
transition points, ordered from most to least transition. Figure~\ref{fig.hardness_plots} plots the results for test set $D_1$ across the baseline and the SFN+C model, showing that our SFN+C model performs better than the baseline at all rankings. Further, the biggest improvements over the baseline occur at instances of greatest transition, a finding that we also discover for other test sets and models. Compared to the baseline, the proposed models are more helpful when forecasting users' sudden engagement shifts.

A subset of transition points are sequences where a user shifts from some news engagement to no news engagement at all or vice versa. These shifts likely suggest that the user is becoming engaged or disengaged with news, which we expect to occur during notable events (e.g., elections and political controversies). In these instances, some users become swept up in the issues, while others avoid the flood of information. We measure the performance of these methods by first selecting all samples where there is some news engagement activity in a particular input step but no news engagement in the output step, or where there is no news engagement in the input step but news engagement in the output step. The results in Table \ref{tab:trans_perf} show improvements over the baseline for nearly all datasets.
Notably, the SFN+T model outperforms both the SFN+C model and the baseline for $D_3$ and $D_4$ when predicting a no engagement-to-engagement shift. We surmise that text features provide a key signal for users' increasing engagement with news topics. 

\begin{table}[ht]
% \small
\centering
\begin{tabular}{cccc}
\hline
\textbf{Dataset}    & \textbf{Models} & \textbf{\begin{tabular}[c]{@{}c@{}}$E \rightarrow N$\\ MAE\end{tabular}} & \textbf{\begin{tabular}[c]{@{}c@{}}$N \rightarrow E$\\ MAE\end{tabular}} \\ \hline
\multirow{4}{*}{$D_1$} & Baseline        & 0.653                                                                        & \textbf{0.818}                                                                        \\
                    & SFN+C           & 0.558                                                                        & 0.819                                                                        \\
                    & SFN+T           & 0.660                                                                        & 0.828                                                                        \\
                    & MFN             & \textbf{0.549}                                                                        & 0.842                                                                        \\ \hline
\multirow{4}{*}{$D_2$} & Baseline        & 0.651                                                                        & 0.930                                                                        \\
                    & SFN+C           & \textbf{0.503}                                                                        & \textbf{0.921}                                                                        \\
                    & SFN+T           & 0.674                                                                        & 0.958                                                                        \\
                    & MFN             & 0.542                                                                        & 0.938                                                                        \\ \hline
\multirow{4}{*}{$D_3$} & Baseline        & 1.327                                                                        & 2.144                                                                        \\
                    & SFN+C           & \textbf{1.016}                                                                        & 2.121                                                                        \\
                    & SFN+T           & 1.267                                                                        & \textbf{2.120}                                                                        \\
                    & MFN             & 1.029                                                                        & 2.139                                                                        \\ \hline
\multirow{4}{*}{$D_4$} & Baseline        & 1.307                                                                        & 2.535                                                                        \\
                    & SFN+C           & \textbf{1.268}                                                                        & 2.496                                                                        \\
                    & SFN+T           & 1.566                                                                        & 2.485                                                                        \\
                    & MFN             & 1.561                                                                        & \textbf{2.474}                                                                        \\ \hline
\end{tabular}
\caption{Mean absolute error for instances where users shift from Engagement to No Engagement ($E \rightarrow N$) and from No Engagement to Engagement ($N \rightarrow E$).}
\label{tab:trans_perf}
\end{table}

\subsection{Text Analysis}
To examine text features in more detail, we identify terms that are leading indicators of engagement with unreliable news sources (-3 or +3). Identifying salient terms from LSTMs is challenging, and we perform a number of steps to identify and analyze terms that are associated with unrealiable news. First, we rank all instances by the model's forecasted engagement counts for news sources with -3 and +3 news content. We then select the top 300 and bottom 300 instances, i.e., those times when users are forecast to most-engage and least-engage with unreliable content. To identify the terms that distinguish these two sets of users, for each user, we concatenate the terms from all of the tweets in the input window and perform a chi-square analysis to select the most distinguishing terms for each user group. The most representative terms used in the context of unreliable liberal (-3) and unreliable conservative (+3) news sources are presented in Tables \ref{tab:libs_fake_top_terms} and \ref{tab:cons_fake_top_terms}, respectively.

We observe that there are differences in terms of focus and referencing method. For example, content that precedes engagement with unreliable liberal sources highlights COVID-19, specific politicians, specific policies (e.g., health policy, tax policy, gun policy), Trump's impeachment, the distinction between truth and lies, the private sector, racism, Russia and Putin, and Muslims. For conservative sources, the content focuses extensively on specific groups with a potential role to play within American political institutions, including globalists, Marxists, Antifa, and communists. Beyond those groups, there is also reference to groups that could be based within or outside the United States, including illegal immigrants, Hezbollah, Islamists, traffickers (drug and migrant), the drug cartel, and individuals affiliated with Hezbollah, Venezuela, Iranians, Libya, and jihad.

\begin{table}[t]
    \begin{subtable}[t]{.49\textwidth}
    \small
    \centering
    \begin{tabular}{|p{.9\textwidth}|}
    \hline
    policy, senators, jail, director, test, muslim, lie, 2016, leaders, attack, technologies, services,  dead, administration, coronavirus, won, senator, centre, sector, role, democracy, biden, donald,  russia, billion, justice, impeached, obama, fake, general, tax, limited, army, journey, yesterday, fucking, various, truth, rights, sir, companies, bank, votes, democratic, biggest, officials, address,   protect, intelligence, hospital, asking, deaths,   congratulations, startups, impeachment, political,   cristiano, nra, military, supporters, senate,   voters, massive, healthcare, january, total, guy,   trump, chinese, growth, russian, cases, digital, racist,  technology, court, muslims, putin, america, sen, hell,   \#iot, successful, election, fox, investigation, lying,   press, graham, rep, ronaldo, texas, republican, leader,   ahead, gop, congress\\
    \hline
    \end{tabular}
    \caption{}
    \label{tab:libs_fake_top_terms}
    \end{subtable}
    \begin{subtable}[t]{.49\textwidth}
    \small
    \centering
    \begin{tabular}{|p{.9\textwidth}|}
    \hline
    globalist, lefts, marxist, illegals, islam, hamas,   marxists, leftists, soros, msm, censoring, veritas,    sharia, globalists, \#antifa, ccp, farleft,   aborted, cartels, noncitizens, hezbollah, venezuela,   islamist, declassified, amnesty, traffickers, fisa,   iranians, cartel, deletes, levin, spied, rino,   indoctrination, libya, marxism, jihad, islamic,   smuggling, passports, declassify, bombshell,   leftwing, lid, russiagate, durham, spying, \#walkaway,   strzok, censors, dominion, leftist, erupts, biological,   harvesting, unborn, communists, communism, bribes,  behar, stabbed, omars, satanic, nadler, sanctuary,   rashida, flashback, pcr, alqaeda, brotherhood,   censorship, totalitarian, \#foxnews, wuhan, assange,   clapper, antitrump, accusers, informant,   rinos, censored, maxine, bureaucrats, deception,   explodes, lefties, destroys, cabal, irans,   parenthood, accuser, infanticide, kerry,   dossier, faucis, jabs, lawabiding, sharpton\\
    \hline
    \end{tabular}
    \caption{}
    \label{tab:cons_fake_top_terms}
    \end{subtable}
    \caption{Terms that are leading indicators of engagement with unreliable (a) liberal and (b) conservative news sources.}
\end{table}

\subsection{Temporal Cluster Analysis} \label{subsec:cluster_analysis}

Turning now to the identification of salient patterns of news engagement over time, rather than simply analyzing partisan scores over the 2018-2021 period, we integrate tweet content to discover more nuanced engagement patterns. We build upon prior work that uses the latent representation learned by a neural network as the input to a clustering algorithm~\cite{huang2014deep,xu2015short}. To do so, we use the MFN model to represent each user. Specifically, for each prediction window for user $i$, we compute the final network layer $g_{inter}^i$ (Equation~\ref{equ:bi_lstm_mfn_out}), which is the model's representation of all of the content and engagement features from the prior two years. Regarding the user's representation over time, we concatenate the $g_{inter}^i$ vector for each of the 16 prediction windows from 2018-2021 (four per year), resulting in a single vector for each user.

Salient user clusters are determined by applying $k$-means clustering (K=20) to these user vectors.\footnote{We experimented with other values of K, which produced similar results. We settled on 20 given that it offered a good tradeoff between cluster cohesion and the number of clusters.} 

\begin{figure*}[ht]
    \centering
    \includegraphics[width=.95\linewidth]{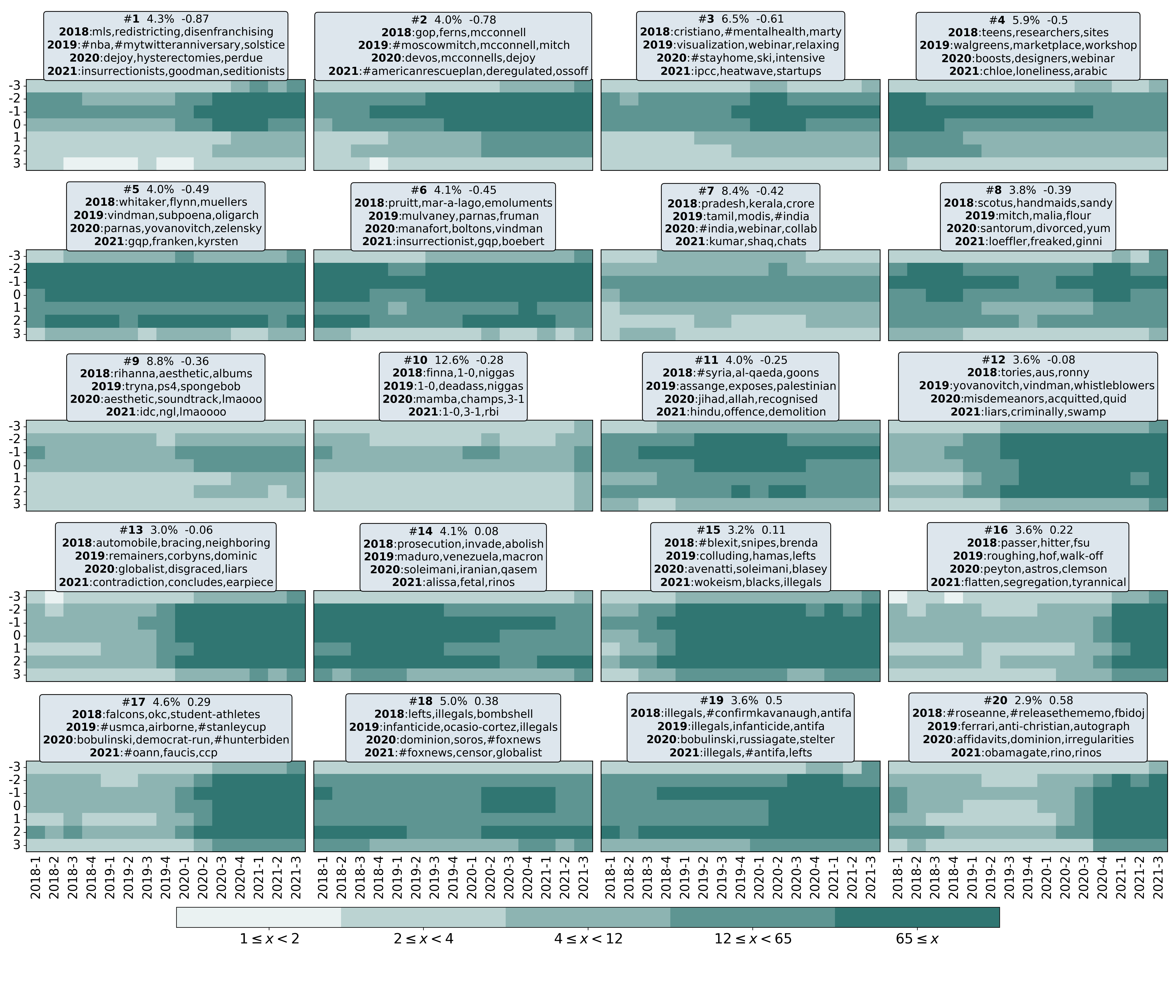}
    \caption{Visualization of 20 discovered clusters of users based on the learned representation of the forecasting model. Over time and partisan stance, each heatmap/cluster includes: percentage of all users within the cluster, the average partisan lean of users in the cluster, and the most distinctive terms for each year.}
    \label{fig.user_clusters}
\end{figure*}

Figure \ref{fig.user_clusters} visualizes both the news engagement and terms used in each cluster. Each heat map displays the number of news engagements per partisan stance in each time window, averaged over the users in the cluster. These counts are discretized into five bins for visualization. The content above each cluster contains the cluster number, the percentage of users belonging to the cluster, the average stance of news engagements in that cluster, and the top-three terms most indicative of the cluster for each year. The terms are ranked by calculating chi-squared statistics for terms appearing in one cluster versus any other cluster. The clusters are presented in the order of their average stance, from most liberal to most conservative.

As an example, we observe that cluster \#1 contains 4.3\% of the users. The average partisan lean of news engagements in this cluster is -.87, which is the most liberal cluster identified. The heatmap indicates that users in this cluster primarily engage with partisan stances -2, -1, and 0. We can also see that users were somewhat engaged early in 2018, but then had less engagement until a spike beginning in late 2020. Looking at the most distinctive terms for each year, we note that much of 2018-2019 was focused on sports (``mls''=Major League Soccer, \#nba=National Basketball Association), with few political term (``disenfranchising,'' ``redistricting''). From 2020, however, the language became more focused on politics: ``hysterectomies'' refers to a news report of detained immigrants being pressured into gynecological procedures; ``DeJoy'' was the U.S. Postmaster General, who was involved in a controversy about changes in mail delivery that impacted mail-in voting, ``Perdue'' refers to a close Senate election in Georgia. In 2021, all three top words refer to the January 6 attack on the U.S. Capitol.

At the other end of the political spectrum, cluster \#17 also contains users who were not at first very politically engaged. Keywords focused initially on sports (``Falcons'' (football) and \#stanleycup (hockey)); yet, in 2020, users became engaged with topics such as Hunter Biden, son of future President Joe Biden, along with his business partner Tony Bobulinski, both of whom were accused of corruption by the Trump campaign. By 2021, these users were focused on the government's response to the pandemic (Fauci), the conservative news network OANN, and tensions between the U.S. and China (``ccp''=Chinese Communist Party).

By using the LSTM representation, we are also able to distinguish between clusters that have similar partisan engagement patterns but discuss different topics. For example, clusters \#18 and \#19 are both strongly conservative with some variation regarding when users first become politically engaged. While both clusters focus on immigration (``illegals'') and abortion (``infanticide''), \#18 is more focused on George Soros, \#19 is more focused on far-left groups (Antifa) and investigations into Biden (``bobulinski,'' ``russiagate'').

There are two intuitive findings from this visualization. First, users who engage with less hyperpartisan news also have lower engagement overall. For example, clusters \#7, \#9, \#10 have the lowest engagement and also tend to engage mostly with -1 and 0 partisan stances. This is in line with prior work suggesting that the emotional language of hyperpartisan news encourages greater engagement online~\cite{hasell2021shared,eady2021,weismueller2022makes}. Second, engagement with contra-partisan news varies across the political spectrum. For example, comparing liberal clusters \#1-\#6 with conservative clusters \#14-\#20, we observe that liberal clusters have substantially less engagement with conservative stances than conservative clusters have with liberal stances. Cluster \#19, for example, engages heavily with both -1 and +3 sources, while cluster \#3 engages little with conservative sources. This finding has additional implications given that \citenoun{bail2018exposure} found that exposure to opposing views on social media can increase political polarization. This is also indicative of animosity towards cross-partisan news and politicians, which has been implicated in misinformation sharing and filter bubbles~\cite{osmundsen2021partysharenews,rathje2021out}.

While follow-up studies are required to more rigorously investigate these and other hypotheses, this analysis provides insights into the patterns and prevalence of long-term news engagement behaviors.

\section{Discussion and Limitations}

We have offered a methodology for studying how users engage with political news over long time periods. The results suggest that the future news engagement behaviors of users can be predicted reasonably well based on prior user behavior. However, sudden shifts in behavior are still difficult to predict, and these can often occur due to external political events (e.g., elections, protests, etc.). Clustering and visualizing users based on the latent representation learned by the forecasting model can serve as an exploratory data analysis method to motivate future work in this area.

\subsection{Limitations}
There are several important limitations to this work to consider.  First, by design, our dataset focuses on users with high news engagement (although we also included a random sample of Twitter users). Of course, such users are quite different from the wider population, and so one should be cautious when generalizing our results beyond the sample. Second, we have not attempted to understand the \textit{intent} behind each news engagement. As discussed in the previous section, many users engage with cross-partisan news sources to ridicule rather than support them. Distinguishing among these cases will be essential for future work on this topic, e.g., by using techniques from \citenoun{shivaram2024characterizing}. Third, it is quite possible that the text-based models exhibit varying error rates across demographic groups, though we have not empirically investigated this. As Table~\ref{tab:performance_by_stance} shows, error rates vary by partisan stance; hence, forecast quality could vary by a user's demographic characteristics just as it varies by a user's political preferences.

\subsection{Ethics and Broader Impacts}
\begin{itemize}
    \item \textbf{Negative societal impacts:} While our goal is to understand extant user behavior, forecasting a user's future news engagements can be viewed as a type of user profiling. This may present societal risks if such a technology is used to, for example, censor users by political ideology or conduct targeted advertising to widen partisan divisions. Similarly, it could be used to target users susceptible to misinformation sharing.
    \item \textbf{Cost of misclassification:} There is currently limited cost to the errors of the proposed approach, but if such a model were to be incorporated into a process to restrict the spread of misinformation, then forecasting -3/+3 engagements erroneously could lead to over-aggressive content filtering.
    \item \textbf{Privacy and consent}: We have released an anonymized dataset where each record contains the news source, partisan lean, day, and a unique, anonymous user ID. To comply with terms of service, no raw tweets will be shared. All data is derived from publicly available sources and no interventions are performed. As such, the study was determined to be exempt by the institution's IRB committee. While we have done our best to anonymize the data, it is conceivable that one could guess some of the user names, by cross-referencing the sequence of news sources they engaged with over time.
    \item \textbf{FAIR:} Our data is  released as a simple CSV file with full documentation. It is thus  findable (via link from the paper on the AAAI/ICWSM website), accessible (CSV is an open format), interoperable (CSV can be opened by any system), re-usable (future research on news engagement is possible with the data).
\end{itemize}

\section*{Acknowledgements}
This work was supported by the National Science Foundation Award \#1927407. AC was funded in part by the Tulane's Jurist Center for Artificial Intelligence and by Tulane's Center for Community-Engaged Artificial Intelligence.

\bibliography{references.bib}

@inproceedings{shivaram2024characterizing,
  title={Characterizing Online Criticism of Partisan News Media Using Weakly Supervised Learning},
  author={Shivaram, Karthik and  Bilgic, Mustafa and  Shapiro, Matthew A and Culotta, Aron},
  booktitle={Proceedings of the International Conference on Web and Social Media},
  year={2024}
}

@inproceedings{tsur2015frame,
  title={A frame of mind: Using statistical models for detection of framing and agenda setting campaigns},
  author={Tsur, Oren and Calacci, Dan and Lazer, David},
  booktitle={Proceedings of the 53rd Annual Meeting of the Association for Computational Linguistics and the 7th International Joint Conference on Natural Language Processing (Volume 1: Long Papers)},
  pages={1629--1638},
  year={2015}
}

@inproceedings{islam2023weakly,
  title={Weakly Supervised Learning for Analyzing Political Campaigns on Facebook},
  author={Islam, Tunazzina and Roy, Shamik and Goldwasser, Dan},
  booktitle={Proceedings of the International AAAI Conference on Web and Social Media},
  volume={17},
  pages={411--422},
  year={2023}
}

@article{liu2022climate,
  title={“Climate change” vs.“global warming”: A corpus-assisted discourse analysis of two popular terms in The New York Times},
  author={Liu, Ming and Huang, Jingyi},
  journal={Journal of World Languages},
  volume={8},
  number={1},
  pages={34--55},
  year={2022},
  publisher={De Gruyter}
}

@inproceedings{oreshkin2019n,
  title={N-BEATS: Neural basis expansion analysis for interpretable time series forecasting},
  author={Oreshkin, Boris N and Carpov, Dmitri and Chapados, Nicolas and Bengio, Yoshua},
  booktitle={International Conference on Learning Representations},
  year={2019}
}

@inproceedings{zeng2023transformers,
  title={Are transformers effective for time series forecasting?},
  author={Zeng, Ailing and Chen, Muxi and Zhang, Lei and Xu, Qiang},
  booktitle={Proceedings of the AAAI conference on artificial intelligence},
  volume={37},
  pages={11121--11128},
  year={2023}
}

@article{hasell2021shared,
  title={Shared emotion: The social amplification of partisan news on Twitter},
  author={Hasell, Ariel},
  journal={Digital Journalism},
  volume={9},
  number={8},
  pages={1085--1102},
  year={2021},
  publisher={Taylor \& Francis}
}

@article{weismueller2022makes,
  title={What makes people share political content on social media? The role of emotion, authority and ideology},
  author={Weismueller, Jason and Harrigan, Paul and Coussement, Kristof and Tessitore, Tina},
  journal={Computers in Human Behavior},
  volume={129},
  pages={107150},
  year={2022},
  publisher={Elsevier}
}

@inproceedings{xu2015short,
  title={Short text clustering via convolutional neural networks},
  author={Xu, Jiaming and Wang, Peng and Tian, Guanhua and Xu, Bo and Zhao, Jun and Wang, Fangyuan and Hao, Hongwei},
  booktitle={Proceedings of the 1st Workshop on Vector Space Modeling for Natural Language Processing},
  pages={62--69},
  year={2015}
}

@inproceedings{huang2014deep,
  title={Deep embedding network for clustering},
  author={Huang, Peihao and Huang, Yan and Wang, Wei and Wang, Liang},
  booktitle={2014 22nd International conference on pattern recognition},
  pages={1532--1537},
  year={2014},
  organization={IEEE}
}

@article{wischnewski2021shareworthiness,
  title={Shareworthiness and Motivated Reasoning in Hyper-Partisan News Sharing Behavior on Twitter},
  author={Wischnewski, Magdalena and Bruns, Axel and Keller, Tobias},
  journal={Digital Journalism},
  pages={1--23},
  year={2021},
  publisher={Taylor \& Francis}
}

@article{brady_etal2017,
author = {William J. Brady  and Julian A. Wills  and John T. Jost  and Joshua A. Tucker  and Jay J. Van Bavel },
title = {Emotion shapes the diffusion of moralized content in social networks},
journal = {Proceedings of the National Academy of Sciences},
volume = {114},
number = {28},
pages = {7313-7318},
year = {2017},
doi = {10.1073/pnas.1618923114},
URL = {https://www.pnas.org/doi/abs/10.1073/pnas.1618923114},
eprint = {https://www.pnas.org/doi/pdf/10.1073/pnas.1618923114},
}

@article{valenzuela2017,
    author = {Valenzuela, Sebastián and Piña, Martina and Ramírez, Josefina},
    title = "{Behavioral Effects of Framing on Social Media Users: How Conflict, Economic, Human Interest, and Morality Frames Drive News Sharing}",
    journal = {Journal of Communication},
    volume = {67},
    number = {5},
    pages = {803-826},
    year = {2017},
    month = {08},
    issn = {0021-9916},
    doi = {10.1111/jcom.12325},
    url = {https://doi.org/10.1111/jcom.12325},
    eprint = {https://academic.oup.com/joc/article-pdf/67/5/803/22484141/jjnlcom0803.pdf},
}

@article{van2020antecedents,
  title={Antecedents and consequences of COVID-19 conspiracy theories: a rapid review of the evidence},
  author={van Mulukom, Valerie and Pummerer, Lotte and Alper, Sinan and Cavojova, Vladimira and Farias, J{\'e}ssica Esther Machado and Kay, Cameron Stuart and Lazarevic, Lili and Lobato, Emilio Jon Christopher and Marinthe, Ga{\"e}lle and Banai, Irena Pavela and Šrol, Jakub  and Žeželj, Iris},
  year={2020},
  journal={PsyArXiv}
}

@article{robertson2021engagement,
  title={Engagement Outweighs Exposure to Partisan and Unreliable News within Google Search},
  author={Robertson, Ronald E and Green, Jon and Ruck, Damian and Ognyanova, Katya and Wilson, Christo and Lazer, David},
  journal={arXiv preprint arXiv:2201.00074},
  year={2021}
}

@article {nyhan2021,
	author = {Nyhan, Brendan},
	title = {Why the backfire effect does not explain the durability of political misperceptions},
	volume = {118},
	number = {15},
	year = {2021},
	doi = {10.1073/pnas.1912440117},
	publisher = {National Academy of Sciences},
	issn = {0027-8424},
	URL = {https://www.pnas.org/content/118/15/e1912440117},
	eprint = {https://www.pnas.org/content/118/15/e1912440117.full.pdf},
	journal = {Proceedings of the National Academy of Sciences}
}

@article{aslett2022,
author = {Kevin Aslett  and Andrew M. Guess  and Richard Bonneau  and Jonathan Nagler  and Joshua A. Tucker },
title = {News credibility labels have limited average effects on news diet quality and fail to reduce misperceptions},
journal = {Science Advances},
volume = {8},
number = {18},
pages = {eabl3844},
year = {2022},
doi = {10.1126/sciadv.abl3844},
URL = {https://www.science.org/doi/abs/10.1126/sciadv.abl3844},
eprint = {https://www.science.org/doi/pdf/10.1126/sciadv.abl3844}}

@article{pennycook2021shiftingattention,
author = {Pennycook, Gordon and Epstein, Ziv and Mosleh, Mohsen and Arechar, Antonio A. and Eckles, Dean and Rand, David G.},
title = {Shifting attention to accuracy can reduce misinformation online},
journal = {Nature},
volume = {592},
number = {7855},
pages = {590-595},
year = {2021},
doi = {10.1038/s41586-021-03344-2},
URL = {https://doi.org/10.1038/s41586-021-03344-2},
}

@article{bhuiyan2021nudgecred,
  title={NudgeCred: Supporting News Credibility Assessment on Social Media Through Nudges},
  author={Bhuiyan, Md Momen and Horning, Michael and Lee, Sang Won and Mitra, Tanushree},
  journal={Proceedings of the ACM on Human-Computer Interaction},
  volume={5},
  number={CSCW2},
  pages={1--30},
  year={2021},
  publisher={ACM New York, NY, USA}
}

@article{bak2022combining,
  title={Combining interventions to reduce the spread of viral misinformation},
  author={Bak-Coleman, Joseph B and Kennedy, Ian and Wack, Morgan and Beers, Andrew and Schafer, Joseph S and Spiro, Emma S and Starbird, Kate and West, Jevin D},
  journal={Nature Human Behaviour},
  pages={1--9},
  year={2022},
  publisher={Nature Publishing Group}
}

@inproceedings{introne2020mapping,
  title={Mapping the narrative ecosystem of conspiracy theories in online anti-vaccination discussions},
  author={Introne, Joshua and Korsunska, Ania and Krsova, Leni and Zhang, Zefeng},
  booktitle={International Conference on Social Media and Society},
  pages={184--192},
  year={2020}
}

@article{haber2021research,
  title={Research note: Lies and presidential debates: How political misinformation spread across media streams during the 2020 election},
  author={Haber, Jaren and Singh, Lisa and Budak, Ceren and Pasek, Josh and Balan, Meena and Callahan, Ryan and Churchill, Rob and Herren, Brandon and Kawintiranon, Kornraphop},
  journal={Harvard Kennedy School Misinformation Review},
  year={2021}
}

@article{bail2018exposure,
  title={Exposure to opposing views on social media can increase political polarization},
  author={Bail, Christopher A and Argyle, Lisa P and Brown, Taylor W and Bumpus, John P and Chen, Haohan and Hunzaker, MB Fallin and Lee, Jaemin and Mann, Marcus and Merhout, Friedolin and Volfovsky, Alexander},
  journal={Proceedings of the National Academy of Sciences},
  volume={115},
  number={37},
  pages={9216--9221},
  year={2018},
  publisher={National Acad Sciences}
}

@article{osmundsen2021partisan,
  title={Partisan polarization is the primary psychological motivation behind political fake news sharing on Twitter},
  author={Osmundsen, Mathias and Bor, Alexander and Vahlstrup, Peter Bjerregaard and Bechmann, Anja and Petersen, Michael Bang},
  journal={American Political Science Review},
  volume={115},
  number={3},
  pages={999--1015},
  year={2021},
  publisher={Cambridge University Press}
}

@article{guess2021consequences,
  title={The consequences of online partisan media},
  author={Guess, Andrew M and Barber{\'a}, Pablo and Munzert, Simon and Yang, JungHwan},
  journal={Proceedings of the National Academy of Sciences},
  volume={118},
  number={14},
  year={2021},
  publisher={National Acad Sciences}
}

@article{king2017news,
  title={How the news media activate public expression and influence national agendas},
  author={King, Gary and Schneer, Benjamin and White, Ariel},
  journal={Science},
  volume={358},
  number={6364},
  pages={776--780},
  year={2017},
  publisher={American Association for the Advancement of Science}
}

@article{lutzke2019priming,
  title={Priming critical thinking: Simple interventions limit the influence of fake news about climate change on Facebook},
  author={Lutzke, Lauren and Drummond, Caitlin and Slovic, Paul and {\'A}rvai, Joseph},
  journal={Global Environmental Change},
  volume={58},
  pages={101964},
  year={2019},
  publisher={Elsevier}
}

@article{lobato2020factors,
  title={Factors predicting willingness to share COVID-19 misinformation},
  author={Lobato, Emilio JC and Powell, Maia and Padilla, Lace MK and Holbrook, Colin},
  journal={Frontiers in psychology},
  volume={11},
  pages={2413},
  year={2020},
  publisher={Frontiers}
}

@article{bakshy2015exposure,
  title={Exposure to ideologically diverse news and opinion on Facebook},
  author={Bakshy, Eytan and Messing, Solomon and Adamic, Lada A},
  journal={Science},
  volume={348},
  number={6239},
  pages={1130--1132},
  year={2015},
  publisher={American Association for the Advancement of Science}
}

@inproceedings{bhargava2019gobo,
  title={Gobo: A System for Exploring User Control of Invisible Algorithms in Social Media},
  author={Bhargava, Rahul and Chung, Anna and Gaikwad, Neil S and Hope, Alexis and Jen, Dennis and Rubinovitz, Jasmin and Sald{\'\i}as-Fuentes, Bel{\'e}n and Zuckerman, Ethan},
  booktitle={Conference Companion Publication of the 2019 on Computer Supported Cooperative Work and Social Computing},
  pages={151--155},
  year={2019}
}

@article{dellaposta2020pluralistic,
  title={Pluralistic collapse: The “oil spill” model of mass opinion polarization},
  author={DellaPosta, Daniel},
  journal={American Sociological Review},
  volume={85},
  number={3},
  pages={507--536},
  year={2020},
  publisher={SAGE Publications Sage CA: Los Angeles, CA}
}

@inproceedings{ribeiro2020auditing,
  title={Auditing radicalization pathways on YouTube},
  author={Ribeiro, Manoel Horta and Ottoni, Raphael and West, Robert and Almeida, Virg{\'\i}lio AF and Meira Jr, Wagner},
  booktitle={Proceedings of the 2020 conference on fairness, accountability, and transparency},
  pages={131--141},
  year={2020}
}

@inproceedings{liu2021interaction,
  title={The Interaction between Political Typology and Filter Bubbles in News Recommendation Algorithms},
  author={Liu, Ping and Shivaram, Karthik and Culotta, Aron and Shapiro, Matthew A and Bilgic, Mustafa},
  booktitle={Proceedings of the Web Conference 2021},
  pages={3791--3801},
  year={2021}
}

@inproceedings{masrour2020bursting,
  title={Bursting the Filter Bubble: Fairness-Aware Network Link Prediction},
  author={Masrour, Farzan and Wilson, Tyler and Yan, Heng and Tan, Pang-Ning and Esfahanian, Abdol},
  booktitle={Proceedings of the AAAI Conference on Artificial Intelligence},
  volume={34},
  pages={841--848},
  year={2020}
}

@article{cresci2015fame,
  title={Fame for sale: Efficient detection of fake Twitter followers},
  author={Cresci, Stefano and Di Pietro, Roberto and Petrocchi, Marinella and Spognardi, Angelo and Tesconi, Maurizio},
  journal={Decision Support Systems},
  volume={80},
  pages={56--71},
  year={2015},
  publisher={Elsevier}
}

@misc{eady2021,
title = "News Sharing on Social Media: Mapping the Ideology of News Media Content, Citizens, and Politicians",
author = "Gregory Eady and Richard Bonneau and Joshua Tucker and Jonathan Nagler",
year = "2021",
howpublished="OSF Preprints"
}

@article{rathje2021out,
  title={Out-group animosity drives engagement on social media},
  author={Rathje, Steve and Van Bavel, Jay J and van der Linden, Sander},
  journal={Proceedings of the National Academy of Sciences},
  volume={118},
  number={26},
  year={2021},
  publisher={National Acad Sciences}
}

@article{osmundsen2021partysharenews, 
title={Partisan Polarization Is the Primary Psychological Motivation behind Political Fake News Sharing on Twitter},
volume={115}, 
DOI={10.1017/S0003055421000290}, 
number={3}, 
journal={American Political Science Review}, 
publisher={Cambridge University Press}, 
author={Osmundsen, Mathias and Bor, Alexander and Vahlstrup, Peter Bjerregaard and Bechmann, Anja and Petersen, Michael Bang},
year={2021}, 
pages={999–1015}
}

@article{guess2019fake,
  title={Fake news, Facebook ads, and misperceptions},
  author={Guess, Andrew and Lyons, Benjamin and Montgomery, Jacob M and Nyhan, Brendan and Reifler, Jason},
  journal={Democracy Fund},
  year={2019}
}

@inproceedings{yang2018know,
  title={I know you'll be back: Interpretable new user clustering and churn prediction on a mobile social application},
  author={Yang, Carl and Shi, Xiaolin and Jie, Luo and Han, Jiawei},
  booktitle={Proceedings of the 24th ACM SIGKDD International Conference on Knowledge Discovery \& Data Mining},
  pages={914--922},
  year={2018}
}

@inproceedings{liu2019characterizing,
  title={Characterizing and forecasting user engagement with in-app action graph: A case study of snapchat},
  author={Liu, Yozen and Shi, Xiaolin and Pierce, Lucas and Ren, Xiang},
  booktitle={Proceedings of the 25th ACM SIGKDD International Conference on Knowledge Discovery \& Data Mining},
  pages={2023--2031},
  year={2019}
}

@article{zhang2022twhin,
  title={TwHIN-BERT: A Socially-Enriched Pre-trained Language Model for Multilingual Tweet Representations},
  author={Zhang, Xinyang and Malkov, Yury and Florez, Omar and Park, Serim and McWilliams, Brian and Han, Jiawei and El-Kishky, Ahmed},
  journal={arXiv preprint arXiv:2209.07562},
  year={2022}
}

@article{schuster1997bidirectional,
  title={Bidirectional recurrent neural networks},
  author={Schuster, Mike and Paliwal, Kuldip K},
  journal={IEEE transactions on Signal Processing},
  volume={45},
  number={11},
  pages={2673--2681},
  year={1997},
  publisher={Ieee}
}

@article{hasan2018survey,
  title={A survey on real-time event detection from the Twitter data stream},
  author={Hasan, Mahmud and Orgun, Mehmet A and Schwitter, Rolf},
  journal={Journal of Information Science},
  volume={44},
  number={4},
  pages={443--463},
  year={2018},
  publisher={SAGE Publications Sage UK: London, England}
}

@inproceedings{shivaram2022reducing,
  title={Reducing Cross-Topic Political Homogenization in Content-Based News Recommendation},
  author={Shivaram, Karthik and Liu, Ping and Shapiro, Matthew and Bilgic, Mustafa and Culotta, Aron},
  booktitle={Proceedings of the 16th ACM Conference on Recommender Systems},
  pages={220--228},
  year={2022}
}

@article{grinberg2019fake,
  title={Fake news on Twitter during the 2016 US presidential election},
  author={Grinberg, Nir and Joseph, Kenneth and Friedland, Lisa and Swire-Thompson, Briony and Lazer, David},
  journal={Science},
  volume={363},
  number={6425},
  pages={374--378},
  year={2019},
  publisher={American Association for the Advancement of Science}
}

\section*{Ethics Checklist}

\begin{enumerate}
\item For most authors...
\begin{enumerate}
    \item  Would answering this research question advance science without violating social contracts, such as violating privacy norms, perpetuating unfair profiling, exacerbating the socio-economic divide, or implying disrespect to societies or cultures?
    \answerYes{Yes, see Discussion and Limitations.}
  \item Do your main claims in the abstract and introduction accurately reflect the paper's contributions and scope?
    \answerYes{Yes}
   \item Do you clarify how the proposed methodological approach is appropriate for the claims made? 
    \answerYes{Yes, see Introduction.}
   \item Do you clarify what are possible artifacts in the data used, given population-specific distributions?
    \answerYes{Yes, see Discussion and Limitations.}
  \item Did you describe the limitations of your work?
    \answerYes{Yes, see Discussion and Limitations.}
  \item Did you discuss any potential negative societal impacts of your work?
    \answerYes{Yes, see Discussion and Limitations.}
      \item Did you discuss any potential misuse of your work?
    \answerYes{Yes, see Discussion and Limitations.}
    \item Did you describe steps taken to prevent or mitigate potential negative outcomes of the research, such as data and model documentation, data anonymization, responsible release, access control, and the reproducibility of findings?
    \answerYes{Yes, see Discussion and Limitations.}
  \item Have you read the ethics review guidelines and ensured that your paper conforms to them?
    \answerYes{Yes}
\end{enumerate}

\item Additionally, if your study involves hypotheses testing...
\begin{enumerate}
  \item Did you clearly state the assumptions underlying all theoretical results?
    \answerYes{Yes, see Model Accuracy.}
  \item Have you provided justifications for all theoretical results?
    \answerYes{Yes, see Model Accuracy}
  \item Did you discuss competing hypotheses or theories that might challenge or complement your theoretical results?
    \answerYes{Yes, see Discussion and Limitations}
  \item Have you considered alternative mechanisms or explanations that might account for the same outcomes observed in your study?
    \answerYes{Yes, see Discussion and Limitations}
  \item Did you address potential biases or limitations in your theoretical framework?
    \answerYes{Yes, see Discussion and Limitations}
  \item Have you related your theoretical results to the existing literature in social science?
    \answerYes{Yes, see Related Work.}
  \item Did you discuss the implications of your theoretical results for policy, practice, or further research in the social science domain?
    \answerYes{Yes, see Discussion and Limitations}
\end{enumerate}

\item Additionally, if you are including theoretical proofs...
\begin{enumerate}
  \item Did you state the full set of assumptions of all theoretical results?
    \answerNA{NA}
	\item Did you include complete proofs of all theoretical results?
    \answerNA{NA}
\end{enumerate}

\item Additionally, if you ran machine learning experiments...
\begin{enumerate}
  \item Did you include the code, data, and instructions needed to reproduce the main experimental results (either in the supplemental material or as a URL)?
    \answerNo{Yes.}
  \item Did you specify all the training details (e.g., data splits, hyperparameters, how they were chosen)?
    \answerYes{Yes, see Methods and Experimental Settings}
     \item Did you report error bars (e.g., with respect to the random seed after running experiments multiple times)?
    \answerYes{Yes, see Table \ref{tab:performance_by_stance}.}
	\item Did you include the total amount of compute and the type of resources used (e.g., type of GPUs, internal cluster, or cloud provider)?
    \answerYes{Yes, see Experimental Settings.}
     \item Do you justify how the proposed evaluation is sufficient and appropriate to the claims made? 
    \answerYes{Yes, see Results and Analysis}
     \item Do you discuss what is ``the cost`` of misclassification and fault (in)tolerance?
    \answerYes{Yes, see Discussion and Limitations}
  
\end{enumerate}

\item Additionally, if you are using existing assets (e.g., code, data, models) or curating/releasing new assets...
\begin{enumerate}
  \item If your work uses existing assets, did you cite the creators?
    \answerYes{Yes, see Data.}
  \item Did you mention the license of the assets?
    \answerNA{NA}
  \item Did you include any new assets in the supplemental material or as a URL?
    \answerNo{Yes.}
  \item Did you discuss whether and how consent was obtained from people whose data you're using/curating?
    \answerYes{Yes, see Discussion and Limitations}
  \item Did you discuss whether the data you are using/curating contains personally identifiable information or offensive content?
    \answerYes{Yes, see Discussion and Limitations}
\item If you are curating or releasing new datasets, did you discuss how you intend to make your datasets FAIR?
\answerYes{Yes, see Discussion and Limitations}
\item If you are curating or releasing new datasets, did you create a Datasheet for the Dataset? 
\answerNo{Yes.}
\end{enumerate}

\item Additionally, if you used crowdsourcing or conducted research with human subjects...
\begin{enumerate}
  \item Did you include the full text of instructions given to participants and screenshots?
    \answerNA{NA}
  \item Did you describe any potential participant risks, with mentions of Institutional Review Board (IRB) approvals?
    \answerNA{NA}
  \item Did you include the estimated hourly wage paid to participants and the total amount spent on participant compensation?
    \answerNA{NA}
   \item Did you discuss how data is stored, shared, and deidentified?
    \answerNA{NA}
\end{enumerate}

\end{enumerate}

\appendix

% \clearpage

\section*{Technical Appendix}
\subsection*{Bot Heuristics}

We filter out suspected bot accounts from our initial collection of users, as well as those likely to be celebrities or organizations, by using a set of heuristics from the literature for this filtering step~\cite{cresci2015fame}. Specifically, we compare the characteristics of each account with several cut-off values as follows 
\begin{enumerate}
    \item Follower Size ($\leq 1000$)
    \item Following Size ($\leq 1000$)
    \item Daily Tweet Activity ($\leq 10$)
    \item Total Tweets authored during the life of the account ($\geq 1000$ and $\leq 30000$)
\end{enumerate}

\subsection*{Hyperparameters}

We search over the hyperparameters shown in Table \ref{tab:hyperparameters} to train our models, picking the best settings based on mean absolute error on the validation set. 

\begin{table}[ht]
\centering
\begin{tabular}{ll}
\hline
\multicolumn{1}{c}{\textbf{Parameter}} & \textbf{Values}       \\ \hline
Hidden Units - LSTM                         & 32, 64, 128, 256, 512 \\
Hidden Units - Linear                       & 128, 512              \\
Bidirectional                               & True, False           \\
LSTM Layers                                 & 1, 2                  \\
Activation                                  & Sigmoid, Relu         \\
Learning Rate                               & 1e-3, 1e-4, 1e-5      \\
Batch Size                                  & 32, 64, 128           \\
Early Stopping Patience                     & 3, 5                  \\ \hline
\end{tabular}
\caption{Hyperparameter values tuned on validation data.}
\label{tab:hyperparameters}
\end{table}
\end{document}